\documentclass[aps, prl,reprint, superscriptaddress, floatfix,longbibliography]{revtex4-2}

\usepackage[utf8]{inputenc}
\usepackage{graphicx}
\usepackage{dcolumn}
\usepackage{amsfonts}
\usepackage{bm}
\usepackage{subfigure}
\usepackage[cmyk,dvipsnames]{xcolor}
\usepackage{amsmath}

    \definecolor{pacificb}{HTML}{1CA9C9}

\newcommand{\kalmar}{Department of Mathematics and Physics, Linnaeus University, SE-39231 Kalmar, Sweden}

\newcommand{\iceland}{Science Institute and Faculty of Physical Sciences, University of Iceland, VR-III, 107 Reykjav\'{i}k, Iceland}

\begin{document}
\title{
Quantum spin-wave statistics and dimensionality in thermally activated magnetic transitions}

\author{Jacob J. Mankenberg} 
\email{jacob.mankenberg@lnu.se}
\affiliation{\kalmar}

\author{Grzegorz J. Kwiatkowski}
\affiliation{\iceland}

\author{Pavel F. Bessarab}
\email{pavel.bessarab@lnu.se}
\affiliation{\kalmar}
\affiliation{\iceland}

\begin{abstract}
    A quantum correction to the rate of thermally activated transition between magnetic states is formulated within harmonic transition state theory and the Kramers–Langer theory by employing quantum statistics for stable spin-wave excitations around the initial state and the transition state. Quantum spin-wave statistics modifies both parameters of the Arrhenius expression for the transition rate relative to their classical values: the zero-point fluctuations renormalize the energy barrier, while the Bose-Einstein occupation of the spin-wave modes modifies the pre-exponential factor. Comparison of the classical and quantum-statistical pre-exponential factors reveals a systematic error associated with the excessive weight assigned to high-frequency excitations by classical equipartition. This error accumulates over the spin-wave spectrum and becomes increasingly severe with increasing dimensionality of the system. The effect is demonstrated for coherent magnetization reversal, and transitions involving noncollinear magnetic textures, including collapse of a magnetic skyrmion and rupture of a skyrmion tube, where classical and quantum-statistical predictions for the transition rate can differ by many orders of magnitude in three-dimensional systems.
\end{abstract}
\maketitle




The problem of thermal stability of magnetic states was first addressed theoretically in the seminal works of N\'eel and Brown~\cite{neel1949,brown1963} and has since developed into a broad area of magnetism, covering phenomena such as superpamagnetic relaxation~\cite{coffey2012}, thermally activated magnetization reversal via uniform and nonuniform mechanisms~\cite{bode2004,krause2009,bessarab2013size}, nucleation and depinning of magnetic structures~\cite{gaunt1986,lemerle1998}, time-and temperature-dependence of coercivity~\cite{sharrock1994}, and thermal nucleation and annihilation of topological magnetic textures \cite{sonntag2014,oike2016,wild2017,bessarab2018}. Thermal fluctuations can induce transitions between different local energy minima by driving the system over the energy barrier separating them. 
If these events is rare on the time scale of intrinsic magnetization dynamics, the transition rate 
can be evaluated within a quasi-equilibrium statistical approach. The rate is then determined by the probability for the system to reach the transition state -- the statistically least favorable region of configuration space that the system must traverse on its way from the initial state to the final state -- and by the dynamics of crossing this region.

It is important to realize that the transition rate 
is not defined solely by motion along a single reaction coordinate parametrizing the transition. For a system containing many magnetic moments, there is a large number of stable spin-wave excitations at the initial state and transition state. Their statistical weights can make a substantial entropic contribution to the transition rate~\cite{desplat2020}. This contribution is present even for the simplest mechanism of coherent magnetization reversal: although the reaction coordinate describes a transition in which the magnetization remains uniform, nonuniform spin-wave excitations are also available to the system \cite{bocquet2024}.


The spin-wave statistics can be treated in the harmonic approximation, in which the energy of the system is retained to second order in deviations from the initial state minimum and the first-order saddle point (SP) associated with the transition state. Assuming the absence of zero modes in the system, this leads to the familiar Arrhenius expression for the rate that emerges in the harmonic transition state theory (HTST)~\cite{vineyard1957,bessarab2012harmonic} and Kramers-Langer theory~\cite{kramers1940,langer1969,fiedler2012direct,desplat2018,bocquet2024}:
\begin{equation}
    \Gamma=\nu_{\rm dyn}\frac{Z_{\rm SP}}{Z_{\rm m}}e^{-\beta\Delta E},
    \label{eq:rate_h}
\end{equation}
where $\beta=(k_\text{B}T)^{-1}$, $Z_{\rm SP}$ and $Z_{\rm m}$ are the partition functions of stable harmonic modes around the SP and the initial-state minimum, respectively, and $\Delta E=E_{\rm SP}-E_{\rm m}$ is the energy barrier, with $E_{\rm SP}$ and $E_{\rm m}$ the energy at the SP and the minimum. The HTST and Kramers-Langer approaches differ in the treatment of the dynamical factor $\nu_{\rm dyn}$. In the Kramers–Langer formulation, the formal zero-damping limit yields $\nu_\text{dyn}^\text{KL} = k_BT/2\pi\hbar$. The corresponding HTST dynamical factor generally differs from this value by a factor close to unity ~\cite{lobanov2021}. Both approaches remain applicable as long as the equilibration time within the basin of the initial state is shorter than the retention time; this condition can be satisfied well into the low-damping regime~\cite{bocquet2024}. 

In existing formulations of HTST and Kramers-Langer theory for magnetic systems, the fluctuations are treated classically, and the harmonic partition functions are conventionally expressed in terms of the eigenvalues or determinants of the Hessian of the energy at the stationary points~\cite{bessarab2012harmonic,fiedler2012direct,desplat2018,bocquet2024}. An equivalent representation is obtained in terms of the physical spin-wave frequencies. In the harmonic approximation, the modes are independent, so the partition function factorizes into the single-mode contributions, $z_{\rm cl}=k_B T/(\hbar\omega)$, where $\omega$ is the mode frequency. For a system of $N$ interacting magnetic moments, the excitation spectrum contains $N$ stable modes at the minimum and $N-1$ stable modes at the SP, which results in the following expression for the classical rate:
\begin{equation}
\Gamma_{\rm cl}
=
f_{\text{cl}}e^{-\beta\Delta E},
\label{eq:rate_cl}
\end{equation}
where the Arrhenius pre-exponential factor is
\begin{equation}
f_{\text{cl}}
=
\nu_{\rm dyn}
\beta\hbar
\frac{\prod_{i=1}^{N}\omega_i^{\rm m}}
{\prod_{i=2}^{N}\omega_i^{\rm SP}}.
\label{eq:prefactor_cl}
\end{equation}
Here, $\omega_i^{\rm m}$ and $\omega_i^{\rm SP}$ are the frequencies of the stable spin-wave modes at the initial-state minimum and the SP, respectively. 

The advantage of the frequency representation is that it exposes the relevant energy scale of spin-wave modes: classical statistics applies only when $\hbar\omega\ll k_\text{B}T$. At the same time, spin-wave frequencies in common magnetic materials can reach several tens of THz, corresponding to excitation energies well above the thermal energy at room temperature~\cite{mook1973,perring1991}. This defines a characteristic spin-wave temperature scale, $\Theta\equiv\hbar\omega_{\max}/k_\text{B}$, set by the upper edge $\omega_{\max}$of the spin-wave spectrum. For $T<\Theta$, high-frequency spin-wave modes can lie far outside the classical regime even when the transition itself remains thermally activated. Classical rate theories then assign these modes excessive statistical weight, potentially leading to incorrect estimates of the transition rate. The limitations of classical statistics have long been recognized in theoretical descriptions of particle rearrangements, where quantum corrections are routinely incorporated into the rate expression~\cite{flynn1970,wolynes1981,henkelman2006}.

In this work, we formulate a quantum-statistical extension of the rate theory for magnetic systems. For a stationary configuration $\alpha=\text{m},\text{SP}$, the partition function of the stable modes is
\begin{equation}
Z^{\alpha}_{\rm q}
=
\exp\!\left(-\beta E_{\rm ZP}^{\alpha}\right)
\prod_{i}
\frac{1}{1-\exp\!\left(-\beta\hbar\omega_i^{\alpha}\right)},
\end{equation}
where $E_{\rm ZP}^{\alpha}$ is the sum of zero-point energies of stable spin-wave modes around configuration $\alpha$, whose particular form depends on the microscopic quantization of the given magnetic Hamiltonian \cite{Colpa1978,RoldanMolina2015,Doucot2018}. In addition, quantum effects associated with the unstable mode can be included through a correction factor $\kappa_u(T)$, commonly referred to as the Wigner correction~\cite{wigner1932,miller1974}. The resulting rate can be written as:
\begin{equation}
\Gamma_{\rm q}
=
\kappa(T) f_{\rm q}
e^{-\beta\left(\Delta E+\Delta E_{\rm ZP}\right)},
\label{eq:rate_q}
\end{equation}
where $\Delta E_{\rm ZP}=E_{\rm ZP}^{\rm SP}-E_{\rm ZP}^{\rm m}$, and the pre-exponential factor is
\begin{equation}
f_{\rm q}
=
\nu_{\rm dyn}
\frac{
\prod_{i=1}^{N}
\left[1-\exp\!\left(-\beta\hbar\omega_i^{\rm m}\right)\right]
}{
\prod_{i=2}^{N}
\left[1-\exp\!\left(-\beta\hbar\omega_i^{\rm SP}\right)\right]
}.
\label{eq:prefactor_q}
\end{equation}
Quantum spin-wave statistics therefore affects both parameters of the Arrhenius expression for the rate: zero-point fluctuations renormalize the energy barrier, while the Bose-Einstein occupation of the spin-wave modes modifies the pre-exponential factor. 

At sufficiently low temperature, thermally activated escape is expected to compete with quantum tunneling of magnetization~\cite{thomas1996,bokacheva2000,loth2012,chudnovsky1988}. A characteristic crossover temperature is $T_\text{c}=\hbar\omega_\text{u}/(2\pi k_B)$, where $\omega_\text{u}$ is the magnitude of the unstable-mode frequency at the SP~\cite{chudnovsky1988a,vlasov2016,vlasov2017}. The same temperature scale controls the quantum correction associated with the unstable mode~\cite{miller1974}: for $T\gg T_\text{c}$, the Wigner correction factor behaves as $\kappa(T)=1+\mathcal{O}[(T_\text{c}/T)^2]$ and rapidly approaches unity with increasing temperature. By contrast, quantum-statistical corrections associated with the stable spin-wave modes can remain important up to $T\lesssim\Theta$. The regime of particular interest here is therefore $T_\text{c}\lesssim T\lesssim\Theta$, where the transition remains thermally activated while high-frequency spin-wave modes already require a quantum-statistical description. Since the unstable-mode correction is governed by the lower temperature scale $T_\text{c}$ and does not exhibit the dimensional accumulation considered here, we do not discuss it further.

In what follows, we show that the effect of quantum statistics is not merely a small low-temperature correction to the classical rate. The error introduced by classical equipartition accumulates over the excitation spectrum and becomes more pronounced with increasing dimensionality of the system. We expose this mechanism analytically for coherent magnetization reversal and demonstrate its consequences for skyrmion collapse in two dimensions and rupture of a skyrmion tube in three dimensions, where the classical and quantum-statistical rate predictions can differ by many orders of magnitude.

Consider a system of $N$ magnetic moments arranged on a $d$-dimensional cubic lattice with $L$ moments along each side, $N=L^d$. Assuming the moments interact via isotropic nearest-neighbor exchange and a site-independent local potential acting on each magnetic moment, the energy of the system reads:
\begin{equation}
    E = -\frac{J}{2}\sum_{i=1}^{L^d}\sum_{j\in \mathrm{NN}(i)}\vec{s}_i\cdot\vec{s}_j+K\sum_{i=1}^{L^d}U\left(\vec{s}_{i}\right).
    \label{eq:energy1}
\end{equation}
Here, $\vec{s}_i$ is a unit vector along the magnetic moment at site $i$, $J$ is the strength of the exchange interaction, and $K$ is the parameter setting the energy scale of the on-site term $U(\vec s)$. The specific form of $U(\vec s)$ is not important for the following analysis; we only assume that it possesses a minimum $\vec s_{\rm m}$ and a first-order SP $\vec s_{\rm s}$. Since $U\left(\vec{s}\right)$ depends on the orientation of a single magnetic moment, its local curvature at the stationary point $\vec s_\alpha$ is characterized by the two eigenvalues $\epsilon_{\alpha,1}$ and $\epsilon_{\alpha,2}$ of its Hessian, corresponding to the two tangent-space degrees of freedom. At the minimum, both eigenvalues are positive, whereas at the SP $\epsilon_{{\rm s},1}<0<\epsilon_{{\rm s},2}$. We focus on the exchange-dominated regime, in which uniform rotation of the magnetization remains the transition mechanism. Within HTST and Kramers-Langer theory, this requires the uniform configuration $\vec s_i=\vec s_{\rm s}$, $i=1,\ldots,N$, to remain a first-order SP of the full system, which holds for 
\begin{equation}\label{eq:sys_size}
    L<L_c\equiv\pi\sqrt{\frac{J}{K|\epsilon_{\mathrm{s},1}|}}.
\end{equation}
Restricting the system to spatially uniform configurations recovers the usual single-domain result for the Arrhenius pre-exponential factor $f_0 = \gamma K\sqrt{\epsilon_{\mathrm{m},1}\epsilon_{\mathrm{m},2}}/(2\pi\mu)$ \cite{brown1979,coffey2012}.

\begin{figure}
    \includegraphics[width=0.99\linewidth]{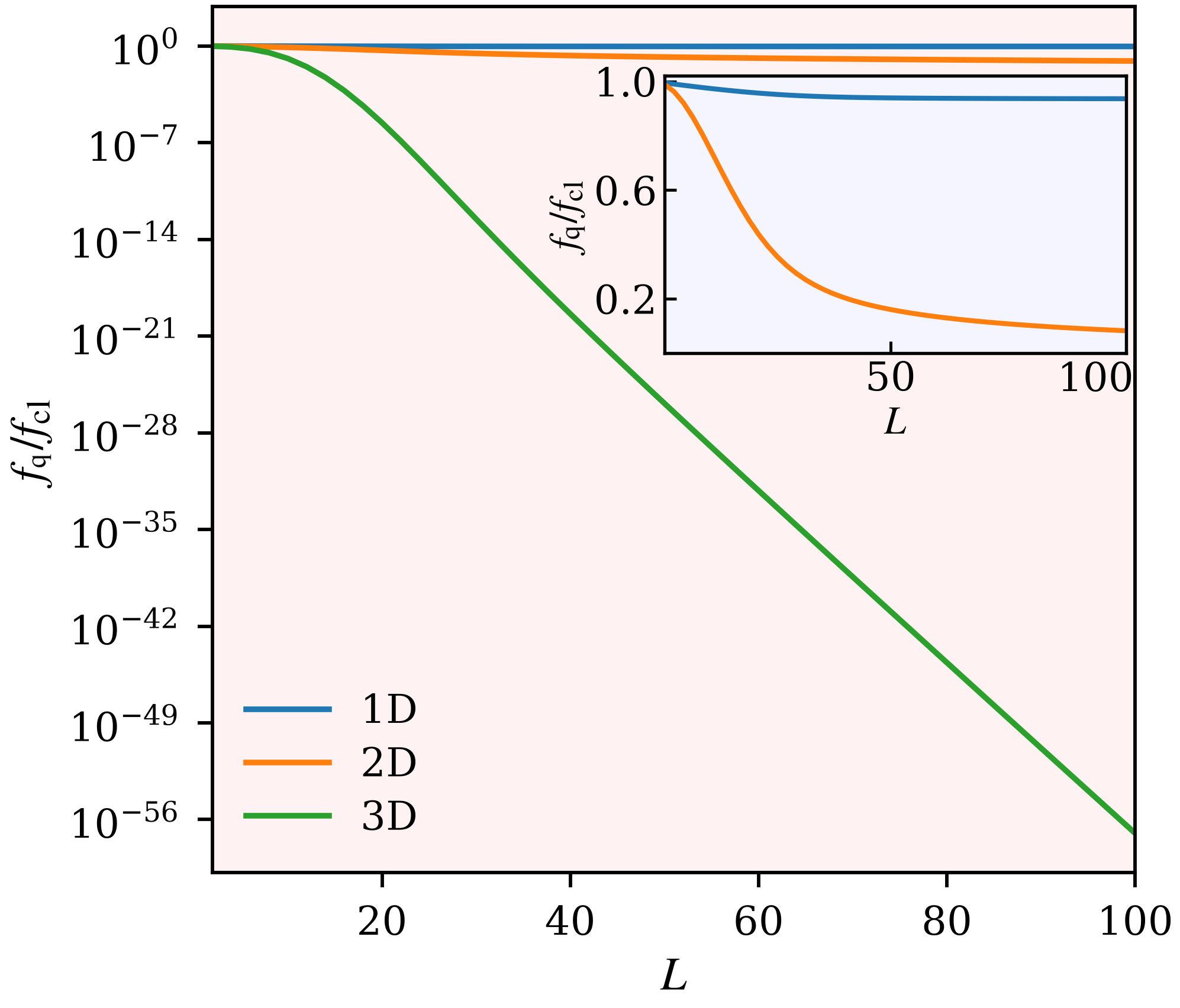}
    \caption{
    Ratio of the quantum-statistical and classical pre-exponential factors as a function of the system size for one-, two-, and three-dimensional spin systems at $\chi=2$ and $T=0.25\Theta$. The ratio $L/L_\text{c}=0.75$ is kept fixed by scaling the exchange interaction with $L$. The inset resolves the curves for one- and two-dimensional systems.}
    \label{fig:prefactor_ratio}
\end{figure}

Despite the uniformity of the transition mechanism in the exchange-dominated regime, nonuniform spin-wave fluctuations are nevertheless available around both the initial-state minimum and the SP. Their contribution is included in both the classical and quantum-statistical expressions for the pre-exponential factor, given by Eqs.~(\ref{eq:prefactor_cl}) and (\ref{eq:prefactor_q}), respectively. Therefore, evaluation of the pre-exponential factor requires the spin-wave spectra at the two stationary configurations. For the Hamiltonian in Eq.~(\ref{eq:energy1}), linearization of the conservative Landau-Lifshitz dynamics about the uniform stationary configurations yields the spin-wave frequencies
\begin{equation}
    \omega_{\alpha}(\mathbf{q}) = \frac{\gamma}{\mu}\sqrt{\left[K\epsilon_{\alpha,1}+J\lambda(\mathbf{q})\right]\left[K\epsilon_{\alpha,2}+J\lambda(\mathbf{q})\right]},
\end{equation}
with
\begin{equation}\label{eq:lambda}
    \lambda(\mathbf{q}) = \sum_{p=1}^{d}4\sin^2\left(\frac{\pi q_p}{2L}\right).
\end{equation}
Here, the index vector $\mathbf{q}=(q_1,\ldots,q_d)$ labels spatial modes, with $q_p=0,\ldots,L-1$.

It is instructive to compare the spin-wave spectra at the energy minimum and at the SP. Defining $\Delta\omega(\mathbf q) \equiv \omega_{\rm m}(\mathbf q)-\omega_{\rm s}(\mathbf q)$ and $\bar{\omega}(\mathbf q)\equiv\left[\omega_{\rm m}(\mathbf q)+\omega_{\rm s}(\mathbf q)\right]/2$, expansion of the mode frequencies to leading order in $K/\left[J\lambda(\mathbf q)\right]$ gives:
\begin{equation}
    \frac{\Delta\omega(\mathbf q)}{\bar{\omega}(\mathbf q)} \approx \frac{K\chi}{2J\lambda(\mathbf q)},
    \label{eq:freq_approx}
\end{equation}
where $\chi=\epsilon_{{\rm m},1}+\epsilon_{{\rm m},2}-\epsilon_{{\rm s},1}-\epsilon_{{\rm s},2}$ characterizes the mismatch between the curvatures of $U(\vec s)$ at the stationary points. In the strong exchange-dominated regime [see Eq.~(\ref{eq:sys_size})], $|\Delta\omega(\mathbf q)|/\bar{\omega}(\mathbf q)\ll 1$ and the spin-wave spectra at the initial-state minimum and the SP become increasingly similar at high frequencies. 

\begin{figure*}
    \centering
    \includegraphics[width=\linewidth]{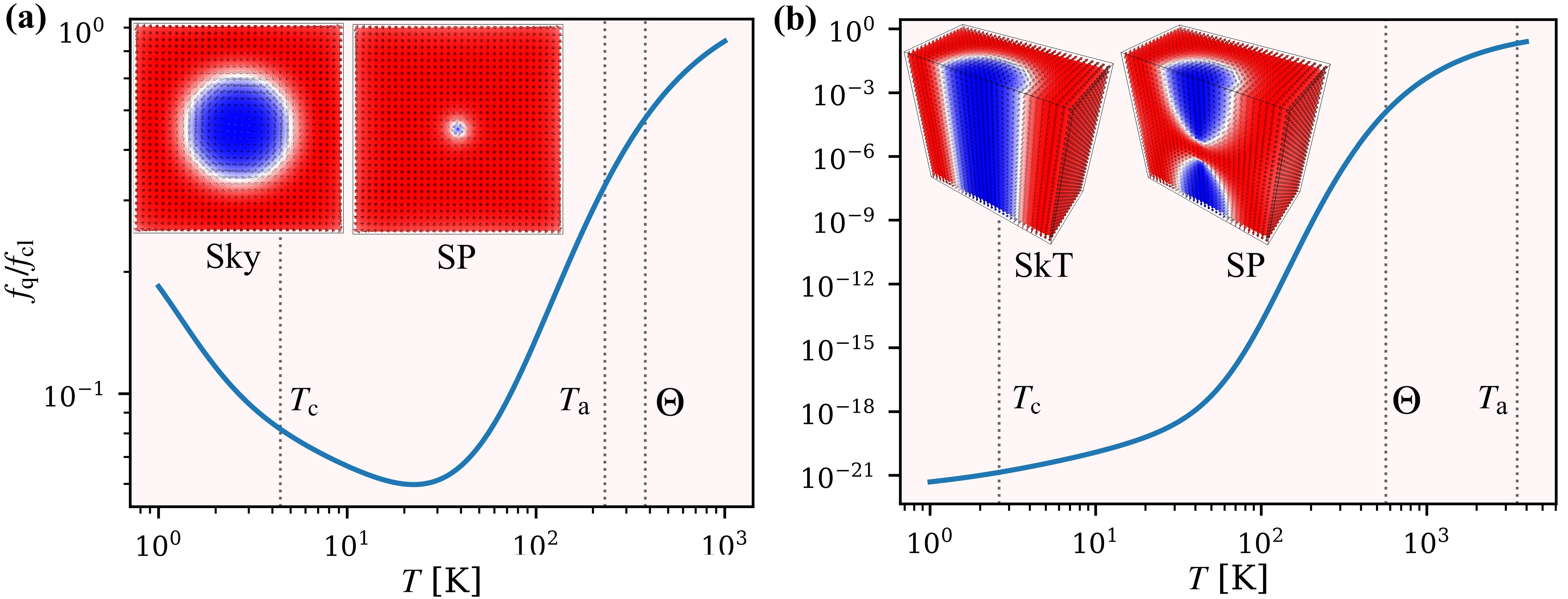}
    \caption{Ratio of the quantum-statistical and classical pre-exponential factors as a function of temperature for (a) the collapse of a two-dimensional skyrmion and (b) the rupture of a three-dimensional skyrmion tube. Insets show the corresponding metastable and SP configurations. The dotted lines mark the tunneling crossover temperature $T_\text{c}$, the characteristic spin-wave temperature scale $\Theta$, and the activation temperature $T_\text{a}=\Delta E/k_\text{B}$, as labeled, with $\Delta E$ being the energy barrier for the corresponding transition.}
    \label{fig:sky_prefactor}
\end{figure*}

Despite the small frequency mismatch, its contribution to the pre-exponential factor accumulates over the entire spin-wave spectrum, and this accumulation can differ strongly between the classical and quantum-statistical descriptions. To quantify this difference, we compare the corresponding pre-exponential factors, Eqs.~(\ref{eq:prefactor_cl}) and (\ref{eq:prefactor_q}), directly. 
Taking the logarithm of the ratio of $f_{\text{q}}/f_{cl}$ converts the products over spin-wave modes into a sum over $\mathbf{q}$, which for sufficiently large $L$ can be replaced by an integral:
\begin{equation}
    \ln\left(\frac{f_{\text{q}}}{f_{cl}}\right) \approx C + \displaystyle\int\limits_{\mathbf{q}\neq\mathbf{0}} d^d\mathbf{q}\bigl[g(\beta\hbar\omega_m) - g(\beta\hbar\omega_s)\bigr],
    \label{eq:prefactor-ratio}
\end{equation}
where $g(x) \equiv \ln[(1-e^{-x})/x]$ and $C = g\bigl(\beta\hbar\omega_{m}(0)\bigr)$. Expanding $g(\beta\hbar\omega)$ around $\bar{\omega}$ gives: 
\begin{equation}
    g(\beta\hbar\omega_m) - g(\beta\hbar\omega_s)\approx-\frac{\Delta\omega}{\bar{\omega}},
    \label{eq:delta-g}
\end{equation}
where $g^\prime(\beta\hbar\omega)\beta\hbar\bar{\omega}\approx -1$ has been used in the low-temperature regime, $\beta\hbar\bar\omega\gg 1$. Substituting Eqs. (\ref{eq:delta-g}) and (\ref{eq:freq_approx}) into Eq. (\ref{eq:prefactor-ratio}) provides a low-temperature estimate of the difference between $f_{cl}$ and $f_{\text{q}}$. To expose how this difference accumulates over the spin-wave spectrum, consider the contribution from modes with $|\mathbf q|\le q_c$. Using $\lambda(\mathbf q)\le \pi^2q^2/L^2$, $q=|\mathbf q|$, and integrating over the corresponding spherical shell in the positive orthant gives the following scaling estimate:
\begin{equation}
    \left.
\ln\left(\frac{f_{\text{q}}}{f_{cl}}\right)
\right|_{q_c}
\approx C - \frac{S_{d-1}}{2^{d+1}}\frac{\chi}{|\epsilon_{s,1}|}\frac{L^2}{L_c^2} \displaystyle\int^{q_c}\limits_{1} dq~q^{d-3},
\label{eq:pref-scaling}
\end{equation}
where $S_{d-1}$ is the surface area of the unit $(d-1)$-sphere. Explicitly, for $d=1,2,$ and $3$:
\begin{equation}
    \left.\ln\left(\frac{f_{\text{q}}}{f_{cl}}\right)
\right|_{q_c} \approx C - \frac{\chi}{4|\epsilon_{s,1}|}\frac{L^2}{L_c^2}
    \begin{cases}
        2(1-q_c^{-1}), & d = 1 \\
        \pi\ln q_c, & d = 2 \\
        \pi(q_c-1), & d = 3
    \end{cases}
\end{equation}
demonstrating that the error introduced into the pre-exponential factor by classical spin-wave statistics becomes increasingly severe with dimensionality. With increasing spectral cutoff $q_c$, it approaches a finite limit in 1D, grows algebraically in 2D, and exponentially in 3D. The convergence in the $d=1$ case is consistent with Braun's continuum treatment of nonuniform magnetization reversal in an effectively one-dimensional ferromagnet, where a finite Arrhenius pre-exponential factor is obtained from the full fluctuation spectrum~\cite{braun1993,braun1994statistical}. In higher dimensions, by contrast, the accumulation of spin-wave modes can produce a difference between the quantum-statistical and classical pre-exponential factors spanning many orders of magnitude. This is illustrated in Fig. \ref{fig:prefactor_ratio}, which shows the ratio $f_{\text{q}}/f_{cl}$ as a function of the system size $L$, calculated directly from Eqs.~(\ref{eq:prefactor_cl}) and (\ref{eq:prefactor_q}) using the full spin-wave spectrum, for $\chi= 4$ and $L/L_c=0.75$ at $T=0.25\Theta$. Therefore, the classical treatment of spin-wave statistics can become inadequate even when the magnetic transition itself remains thermally activated. 
Beyond this analytical demonstration, the quantum-statistical pre-exponential factor is also expected to differ from its classical counterpart for nonuniform magnetic textures. We demonstrate this for the collapse of a two-dimensional skyrmion~\cite{lobanov2016} and the rupture of a three-dimensional skyrmion tube into a pair of chiral bobbers~\cite{rybakov2015}. The calculations are performed for a generic atomistic chiral spin Hamiltonian excluding dipole-dipole interactions~\cite{Muller2019}, with parameters comparable to those employed for atomistic models of three-dimensional skyrmion tubes and chiral bobbers. Specifically, the moments have magnitude $\mu = 1\text{ }\mu_B$, with nearest-neighbor exchange $J=2$ meV, Dzyaloshinskii-Moriya interaction $D=0.55$ meV, easy-axis anisotropy $K=0.10$ meV, and an external field $B=0.1$ T along the easy axis. The two-dimensional and three-dimensional systems contain $32\times32\times1$ and $32\times32\times32$ spins respectively, with open boundary conditions. The global minimum configuration is thus a uniform background with chiral edge twists that lift the translational degeneracy.

For each transition, the initial-state minimum is obtained by relaxing the system from a suitable starting configuration. 
The corresponding first-order SP is then obtained using geodesic minimum-mode following (GMMF)\cite{schrautzer2025identification}. For the rupture of the skyrmion tube, a geodesic nudged elastic band (GNEB)\cite{bessarab2015method} calculation is first used to identify the transition pathway and to generate a configuration sufficiently close to the relevant SP, which is subsequently converged using GMMF. The resulting textures are shown in the insets in Figure~\ref{fig:sky_prefactor}. The SP configuration for the skyrmion collapse is a lattice-scale contracted configuration, while for the skyrmion tube it is a 
configuration involving two nearly touching chiral bobbers. 

The energy barriers for these transitions are $\Delta E^{\rm Sk} = 19.95$ meV for the skyrmion collapse and $\Delta E^{\rm SkT} = 302.65$ meV for the skyrmion tube rupture. The zero-point energy corrections are obtained by rotating the quantum spin Hamiltonian into a local frame of reference aligned with each stationary spin configuration and expanding it in Holstein–Primakoff bosons. For stationary configurations, the linear terms vanish, while the quadratic terms describe the leading quantum fluctuations and determine the zero-point energy \cite{Colpa1978,RoldanMolina2015,Doucot2018}. The resulting corrections to the energy barriers, obtained from the difference between the zero-point energies at the SP and the initial-state minimum, are $\Delta E_{\rm ZP}^{\rm Sk}=-0.07$ meV and $\Delta E_{\rm ZP}^{\rm SkT}=-1.12$ meV. These corrections amount to only 0.35\% and 0.37\% of the classical barriers, respectively. Although the resulting correction to the transition rate increases as $T$ approaches $T_\text{c}$, it remains smaller than the correction associated with the Bose–Einstein occupation of the spin-wave modes. 
%
%
%

Figure~\ref{fig:sky_prefactor} plots the ratio of the quantum-statistical to classical pre-exponential factor for the skyrmion collapse and skyrmion tube rupture.  In both cases, the quantum prefactor approaches the classical result with increasing temperature as progressively higher-frequency spin-wave modes become thermally populated. 
The quantum correction is substantially stronger for the three-dimensional texture, consistent with the enhanced accumulation of high-frequency modes predicted by the dimensional analysis above. 

The applicability of HTST and Kramers–Langer theory requires time of equilibration within the basin of the initial state to remain short compared with the retention time; a useful measure of this condition is $T\ll T_\text{a}$, where $T_\text{a}=\Delta E/k_B$
is the activation temperature set by the energy barrier. At low temperature, thermally activated escape competes with quantum tunneling below the crossover temperature $T_\text{c}$, so the present treatment requires $T\gtrsim T_\text{c}$. Finally, the characteristic spin-wave temperature scale $\Theta$ marks the onset of quantum suppression of high-frequency modes. The relevant temperature regime is therefore $T_\text{c}\lesssim T\lesssim\Theta$, provided that $T\ll T_{\rm a}$. This window is substantial for both transitions shown in Fig.~\ref{fig:sky_prefactor}, where escape remains thermally activated while the spin-wave spectrum already requires a quantum-statistical description. 
This suggests that classical rate theory may substantially underestimate the lifetimes of metastable three-dimensional magnetic textures, including hopfions, for which lifetime calculations have recently been reported~\cite{Sallermann2023,Lobanov2023}.

In conclusion, we have demonstrated within HTST and the Kramers-Langer theory that quantum spin-wave statistics can be crucial for the correct evaluation of the rate of thermally activated magnetic transitions. If the spin-wave spectrum of the system extends beyond the thermal energy scale, the statistical contribution of the fluctuations can no longer be described consistently within classical rate theory. In the exchange-dominated regime, where the high-frequency spin-wave spectra at the initial state and the transition state differ only weakly, the classical treatment introduces only a small error in the entropy difference for each individual pair of modes. However, the cumulative error grows with the size and dimensionality of the system, leading to particularly large discrepancy between the classical and quantum-statistical predictions of the transition rates in three-dimensional systems. The presented approach therefore extends the rate theory of magnetic transitions into an important regime in which thermal activation remains the dominant transition mechanism, while the fluctuations governing the transition probability already require a quantum-statistical description.

In the systems studied here, the dominant quantum-statistical correction to the transition rate enters through the renormalization of the Arrhenius pre-exponential factor by the Bose-Einstein occupation of the spin wave modes. The zero-point energy renormalization of the energy barrier is below $0.4\%$ in both cases and, although its effect on the rate increases as $T$ approaches $T_c$, it remains smaller than the correction to the pre-exponential factor. It remains to be seen whether this hierarchy is generic and how the magnitude of the quantum-statistical correction depends on the topology, morphology, and excitation spectrum of the magnetic structure undergoing the transition.

\begin{acknowledgments}
The authors would like to thank T. Sigurj\'onsd\'ottir, B. ter Hofstede, H. J\'onsson, and V. Uzdin for fruitful discussions. J.J.M. is very grateful to the TWIST group at the Universit\"at Duisburg-Essen, Duisburg, Germany. This work was funded by the Faculty of Technology at Linnaeus University through the strategic research initiative in quantum science and quantum technology and the University Alliance Ruhr. The computational resources were provided by the National Academic Infrastructure for Supercomputing in Sweden (NAISS), funded by the Swedish Research Council.
\end{acknowledgments}

\end{document}